\documentclass[12pt]{article}

\usepackage{latexsym}

  \usepackage{amsfonts}
  \usepackage{amsmath}
\usepackage{amssymb}
\usepackage{amscd}
\usepackage{amsthm}

\usepackage{color}
\usepackage[dvips]{graphicx}

\usepackage{authblk}

\newcommand{\dis}{\displaystyle}
\newcommand{\rf}[1]{(\ref{#1})}
\newcommand{\be}{\begin{equation}}
\newcommand{\ee}{\end{equation}}
\newcommand{\ba}{\begin{array}}
\newcommand{\ea}{\end{array}}

\newcommand{\ods}{\par\vspace{0.2cm}\par}

\newenvironment{Proof}{\par \vspace{2ex} \par
\noindent  {\it Proof:}}{\hfill $\qedsymbol$ 
\vspace{2ex} \par }

\newtheorem{prop}{Proposition}[section]
\newtheorem{Th}[prop]{Theorem}

\newtheorem{cor}[prop]{Corollary}

\numberwithin{equation}{section}

\begin{document}

\title{Fast calculation of inverse square root with the use of magic constant -- analytical approach}

\author[1]{\bf  Leonid V.\,Moroz\thanks{moroz\_lv@polynet.lviv.ua}}
\author[2]{\bf Cezary J. Walczyk\thanks{walcez@gmail.com}}
\author[1]{\bf  Andriy Hrynchyshyn\thanks{hrynchyshyn.a@gmail.com}}
\author[3]{\bf Vijay Holimath\thanks{vijay.holimath@vivid-sparks.com}}
\author[2]{\bf Jan L. Cie\'sli\'nski\thanks{ j.cieslinski@uwb.edu.pl}} 
\affil[1]{\small   Lviv Polytechnic National University, Department of Security Information and Technology, st. Kn. Romana 1/3, 79000 Lviv, Ukraine }
\affil[2]{\small Uniwersytet w Bia\l ymstoku, Wydzia{\l} Fizyki, ul.\ Cio\l kowskiego 1L,  15-245 Bia\l ystok, Poland}
\affil[3]{\small VividSparks IT Solutions, Hubli 580031, No. 38, BSK Layout, India }

\date{}

\maketitle

\abstract{
We present a mathematical analysis of transformations used in fast calculation of  inverse square root for single-precision floating-point numbers. Optimal values of the so called magic constants are derived in a systematic way, minimizing either absolute or relative errors at subsequent stages of the discussed algorithm. 
}
\ods


\section{Introduction}

Floating-point arithmetics has became wide spread in many applications such as 3D graphics, scientific computing and signal processing  \cite{cit12, cit13, cit14}. Basic operators such as addition, subtraction, multiplication are easier to design and yield higher performance, high throughput but advanced operators such as division, square root, inverse square root and trigonometric functions consume more hardware, slower in performance and slower throughput \cite{cit22, cit23, cit24, cit25, cit26}.

Inverse square root function is widely used in 3D graphics especially in lightning reflections \cite{Eberly,cit15, cit16}. Many algorithms can be used to approximate inverse square root functions \cite{cit17, cit18, cit19,cit20, cit21}. All of these algorithms require initial seed to approximate function. If the initial seed is accurate then iteration required for this function is less time-consuming. In other words, the function requires less cycles. In most of the case, initial seed is obtained from Look-Up Table (LUT) and the LUT consume significant silicon area of a chip. In this paper we present initial seed using so called magic constant  \cite{cit1,Lomont} which does not require LUT and we then used this magic constant to approximate inverse square root function using Newton-Raphson method and discussed its analytical approach. 

We present first mathematically  rigorous  description  of the fast algorithm for computing  inverse square root for single-precision IEEE Standard 754 floating-point numbers (type \textbf{float}).

\begin{tabbing}
0000000\=122\=5678\=\kill
\>\textit{1.}\> \textbf{float} InvSqrt(\textbf{float} x)\{ \\
\>\textit{2.}\>\>\textbf{float} halfnumber = 0.5f * x; \\
\>\textit{3.}\>\> \textbf{int} i = *(\textbf{int}*) \&x;\\
\>\textit{4.}\>\> i = R-(i$>>$1);\\
\>\textit{5.}\>\>  x = *(\textbf{float}*)\&i;\\
\>\textit{6.}\>\> x = x*(1.5f-halfnumber*x*x); \\
\>\textit{7.}\>\> x = x*(1.5f-halfnumber*x*x); \\
\>\textit{8.}\>\> \textbf{return} x ;\\
\>\textit{9.}\>\}
\end{tabbing}
This code, written   in \textbf{C},  will be referred to as function \textit{InvSqrt}. It realizes  a fast algorithm for calculation of the inverse square root.  
In line \textit{3} we  transfer bits of varaible  x (type \textbf{float}) to variable  i (type \textbf{int}).  In line \textit{4} we determine an initial value (then subject to the iteration process) of the inverse square root, where  $R=0x5f3759df$  is a ``magic constant''.  In line  \textit{5}  we transfer bits of a variable  i (type \textbf{int}) to the variable  x (type \textbf{float}). Lines  \textit{6} and \textit{7}  contain subsequent iterations of the Newton-Raphson algoritm.

The algorithm  \textit{InvSqrt} has numerous applications, see   \cite{cit3, cit4, cit7, cit8, cit9}.  The most important among them is 3D computer graphics, where  normalization of vectors is ubiquitous.  \textit{InvSqrt}  is characterized by a high speed, more that 3 times higher than in computing the inverse square root using library functions.  This property is discussed in detail in \cite{cit10}.  The errors of the fast inverse square root algorithm depend on the choice of $R$.  In several theoretical papers  \cite{Lomont,cit10, cit5,cit6, cit11} (see also the Eberly's monograph \cite{Eberly})  attempts were made to determine analytically the optimal value  (i.e. minimizing errors) of the magic constant.  These attempts were not fully successfull. In our paper we present  missing  mathematical description of all steps of the fast inverse square root algorithm.  

\section{Preliminaries}
%\section{Single-precision IEEE Standard 754 floating-point numbers}

The value of a floating-point number can be represented as:
\begin{equation}  
x=(-1)^{s_x}(1+m_x)2^{e_x},     \label{r1_1}
\end{equation}
where  $s_x$  is the sign bit ($s_x=1$ for negative numbers and $s_x=0$ for positive numbers), $1+m_x$  is normalized mantissa (or significand), where  $m_x \in \langle 0,\;1)$ and, finally, $e_x$  is an integer. 

\begin{figure}  \label{fig-layout}
\begin{center}
\includegraphics[width=14cm]{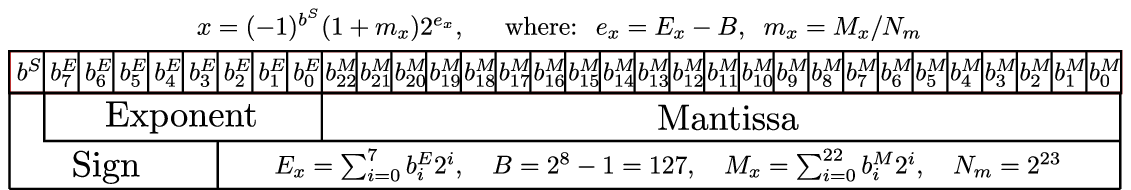}
\end{center}
\caption{The layout of a 32-bit floating-point number.}
\label{pic1}
\end{figure}

In the case of the standard  \textit{IEEE-754} a floating-point number is encoded by $32$ bits (Fig.~\ref{fig-layout}). The first bit corresponds to a sign, next $8$ bits correspond to an exponent  $e_x$  and the last $23$ bits encodes a mantissa.  The fractional part of the mantissa is represented by an integer (without a sign) $M_x$:
\begin{equation}
M_x=N_m\,m_x,\quad\quad \text{where: \  $N_m=2^{23}$},\label{r1_2}  
\end{equation}
and the exponent is represented by a positive value $E_x$  resulting from the shift of  $e_x$ by a constant $B$ (biased exponent):
\begin{equation}
E_x=e_x+B,\quad \quad  \text{where: \  $B=127$}.\label{r1_3}
\end{equation}
Bits of a floating-point number can be interpreted as an integer given by: 
\begin{equation}
I_x=(-1)^{s_x}(N_m\,E_x+M_x),\label{r1_4}
\end{equation}
where:
\begin{equation}
M_x=N_m(2^{-e_x}x-1)\label{r1_5}
\end{equation}

In what follows we confine ourselves to positive numbers ($s_x\equiv b_S=0$). Then, to a given  integer $I_x \in \langle 0, 2^{32} - 1 \rangle$ there corresponds  a floating number $x$ of the form (\ref{r1_1}), where
\begin{equation}  \label{aaa}
e_x := \lfloor N_m^{-1} I_x \rfloor - B \ , \qquad    m_x :=   N_m^{-1} I_x  -  \lfloor N_m^{-1} I_x \rfloor  \ .
\end{equation}
This map, denoted by $f$,  is inverse to the map $x \rightarrow I_x$. In other words, 
\begin{equation}
   f(I_x) = x \  .
\end{equation}

The range of available 32-bit floating-point numbers for which we can determine inverse square roots can be divided into 127 disjoint intervals: 
\begin{equation}
x\in \bigcup_{n=-63}^{63} A_n,\quad\text{where: $A_n=\{2^{2n}\}\cup\underbrace{( 2^{2n},2^{2n+1})}_{A_n^I}\cup\langle \underbrace{2^{2n+1},2^{2(n+1)})}_{A_n^{II}}$}.\label{r2_1}
\end{equation}
Therefore, $e_x = 2 n$ for $x \in A^I_n \cup \{ 2^{2n} \}$ and $e_x = 2 n +1$ for $x\in A^{II}_n$.  
For any $x\in A_n$  exponents and significands of  $y=1/\sqrt{x}$  are given by
\begin{equation}
e_{y}=\left\{
\begin{array}{cl}
-n& \text{for $x=2^{2n}$}\\
-n-1& \text{for $x\in A_n^I$}\\
-n-1& \text{for $x\in A_n^{II}$}
\end{array}
\right.,\quad
m_{y}=\left\{
\begin{array}{cl}
0& \text{for $x=2^{2n}$}\\
2/\sqrt{1+m_{x}}-1& \text{for $x\in A_n^I$}\\
\sqrt{2}/\sqrt{1+m_{x}}-1& \text{for $x\in A_n^{II}$}
\end{array}
\right.     .   \label{r2_2}
\end{equation}
It is convenient to introduce new variables  $\tilde{x}=2^{-2n}x$  and  $\tilde{y}=2^n y$  (in order to have $\tilde{y}=1/\sqrt{\tilde{x}}$). Then:
\begin{equation}
e_{\tilde{y}}=\left\{
\begin{array}{cl}
0& \text{for $x=2^{2n}$}\\
-1& \text{for $x\in A_n^I$}\\
-1& \text{for $x\in A_n^{II}$}
\end{array}
\right.,\quad
m_{\tilde{y}}=\left\{
\begin{array}{cl}
0& \text{for $x=2^{2n}$}\\
2/\sqrt{1+m_{x}}-1& \text{for $x\in A_n^I$}\\
\sqrt{2}/\sqrt{1+m_{x}}-1& \text{for $x\in A_n^{II}$}
\end{array}
\right.    ,     \label{r2_3}
\end{equation}
which means that without loss of the generality we can confine ourselves to $x\in\langle 1,4)$:
\begin{equation}
\left\{
\begin{array}{ll}
\tilde{x}=1& \text{for $x=2^{2n}$}\\
\tilde{x}\in\langle 1,2)& \text{for $x\in A_n^I$}\\
\tilde{x}\in\langle 2,4)& \text{for $x\in A_n^{II}$}
\end{array}
\right.. \label{r2_4}
\end{equation}

%\section{Zeroth approximation of the inverse square root and Newton-Raphston corrections} Inverse square root of a floating-point number and its zeroth approximation
\ods

\section{Theoretical explanation of \textit{InvSqrt} code  }

In this section we present a mathematical interpretation of the code \textit{InvSqrt}. The most important part of the code is contained in the line \textit{4}. Lines \textit{4} and \textit{5}  produce  a zeroth approximation of the inverse square root of  given positive floating-point number $x$ ($s_x=0$). The zeroth approximation will be used as an initial value for the Newton-Raphson iterations (lines \textit{6} and \textit{7}  of the code).

\begin{Th}
The porcedure of determining of an initial value  using the magic constant, described by  lines \textit{4} and \textit{5} of the code,  can be represented by the following function
\begin{equation}  \label{y0}
\tilde{y}_{0} (\tilde x, t)  = \left\{   \begin{array}{ll}   \displaystyle 
-\frac{1}{4} \tilde{x} + \frac{3}{4} + \frac{1}{8} t     &     \text{for}  \ \ \tilde{x} \in \langle 1,2 )    \\[2ex]   \displaystyle 
-\frac{1}{8} \tilde{x} + \frac{1}{2} + \frac{1}{8} t    &  \text{for}  \ \   \tilde{x} \in   \langle 2, t)    \\[2ex]   \displaystyle 
-\frac{1}{16} \tilde{x} + \frac{1}{2} + \frac{1}{16} t      &  \text{for} \ \  \tilde{x} \in  \langle  t,4) 
\end{array}  \right. 
\end{equation}
where 
\be
t = t_x = 2 + 4 m_R + 2 \mu_x N_m^{-1} ,
\ee 
 $m_R :=   N_m^{-1} R  -  \lfloor N_m^{-1} R \rfloor$ and  $\mu_x =0$ for $M_x$ even and  $\mu_x =1$ for $M_x$  odd.  Finally, the floating-point number $f(R)$, corresponding to the magic constant $R$, satisfies
\be
       e_R = 63 \ , \quad   m_R < \frac{1}{2} \ ,
\ee
where $f (R) = (1+m_R) 2^{e_R}$. 
\end{Th}

\begin{Proof}  
The line \textit{4}  in the definition of the  \textit{InvSqrt} function consists of two operations. The first one is a right bit shift of the number $I_x$, defined by (\ref{r1_4}),  which yields the integer part of its half:
\begin{equation}
I_{x/2}=\lfloor I_x/2 \rfloor=2^{-1}N_m(B+e_x)+\lfloor 2^{-1}N_m(2^{-e_x}x-1)\rfloor,      \label{r3_1}
\end{equation}
The second operation yields 
\begin{equation}
I_{y_0}:=R-I_{x/2},\label{r3_2} 
\end{equation}
and $y_0 \equiv f (R-I_{x/2})$ is computed in the line \textit{5}. This floating-point number will be used as a zeroth approximation of the inverse square root, i.e., $y_0\simeq 1/\sqrt{x}$  (for a justification see the next section).   Denoting, as usual, 
\begin{equation}
y_0=(1+m_{y_0})2^{e_{y_0}},\label{r3_3}
\end{equation} 
and remembering that
\be
 R = N_m (e_R + B + m_R) , 
\ee
we see  from \rf{aaa}, \rf{r3_1} and \rf{r3_2} that
\begin{equation}
m_{y_0}=e_R+m_R-e_{y_0}-N_m^{-1} I_{x/2},\quad e_{y_0}=e_R+\lfloor m_R-N_m^{-1}I_{x/2} \rfloor \label{r3_4}.
\end{equation}
Here $e_R$ is an integer part and $m_R$ is a mantissa of the floating-point number given by $f(R)$.  It means that  $e_R=63$ and $m_R<1/2$.  

According to formulas (\ref{r3_1}), (\ref{r3_4}) and (\ref{r1_5}),  confining ourselves to $\tilde{x}\in\langle 1,4)$,  we obtain:
\begin{equation}
I_{\tilde{x}/2}=\lfloor 2^{-1}N_m m_{\tilde{x}}\rfloor+\left\{
\begin{array}{cl}
2^{-1}N_m B & \text{for $\tilde{x}\in\langle 1,2)$}\\
2^{-1}N_m (B+1) & \text{for $\tilde{x}\in\langle 2,4)$}\\
\end{array}
\right. .  \label{r3_5}
\end{equation}
Hence
\begin{equation}
e_{\tilde{y}_0}=e_R-\frac{B+1}{2}+\left\{
\begin{array}{cl}
\lfloor m_R+2^{-1}-N_m^{-1}\lfloor 2^{-1}N_m m_{\tilde{x}}\rfloor\rfloor & \text{for $\tilde{x}\in\langle 1,2)$}\\
\lfloor m_R-N_m^{-1}\lfloor 2^{-1}N_m m_{\tilde{x}}\rfloor\rfloor & \text{for $\tilde{x}\in\langle 2,4)$}\\
\end{array}
\right.. \label{r3_6}
\end{equation}

Therefore, requiring $e_R = \frac{1}{2} (B-1) = 63$  and $m_R < \frac{1}{2}$  we get $e_{\tilde{y}_0} = -1$, which means that $e_{\tilde{y}_0} =e_y$ for $\tilde{x} = \langle 1,2)$. 
The condition  $m_R<1/2$  implies
\begin{equation}
\lfloor m_R+2^{-1}-N_m^{-1}\lfloor 2^{-1}N_m m_{\tilde{x}}\rfloor\rfloor=0,\label{r3_7}
\end{equation}
and
\begin{equation}
\lfloor m_R-N_m^{-1}\lfloor 2^{-1}N_m m_{\tilde{x}}\rfloor\rfloor=\left\{ 
\begin{array}{cl}
0 & \text{for $m_{\tilde{x}}\in\langle 0,2m_R\rangle$}\\
-1 & \text{for $m_{\tilde{x}}\in(2m_R,1)$}
\end{array}
\right.,\label{r3_8}
\end{equation}
which means that 
\begin{equation}
e_{\tilde{y}_0}=\left\{ 
\begin{array}{cl}
e_R-\frac{B+1}{2}=-1 & \text{for $\tilde{x}\in \langle 1,2+4 m_R\rangle$}\\
e_R-\frac{B+3}{2}=-2 & \text{for $\tilde{x}\in(2+4m_R,4)$}
\end{array}
\right. , \label{r3_9}
\end{equation}
which ends the proof.  \end{Proof}

In order to get a simple expression for the mantissa $m_{\tilde{y}_0}$ we can  make a  next approximation: 
\begin{equation}
\lfloor 2^{-1}N_m m_{\tilde{x}}\rfloor\simeq  2^{-1}N_m m_{\tilde{x}}-2^{-1},\label{r3_10}
\end{equation}
which yields a new estimation of the inverse square root:
\begin{equation}
\tilde{y}_{00}=(1+m_{\tilde{y}_{00}})2^{e_{\tilde{y}_{00}}},\label{r3_11}
\end{equation} 
where for  $t=2+4m_R+2N_m^{-1}$:
\begin{equation}
\left\{ 
\begin{array}{lll}
e_{\tilde{y}_{00}}=-1, &   m_{\tilde{y}_{00}}=2^{-2}t-2^{-1}m_{\tilde{x}}&\text{for $\tilde{x}\in \langle 1,2)=\tilde{A}^I$}\\
e_{\tilde{y}_{00}}=-1, & m_{\tilde{y}_{00}}=2^{-2}t-2^{-1}m_{\tilde{x}}-2^{-1}&\text{for $\tilde{x}\in\langle 2,t\rangle=\tilde{A}^{II}$}\\
e_{\tilde{y}_{00}}=-2, & m_{\tilde{y}_{00}}=2^{-2}t-2^{-1}m_{\tilde{x}}+2^{-1}&\text{for $\tilde{x}\in(t,4)=\tilde{A}^{III}$}
\end{array}
\right..\label{r3_12}
\end{equation}

Because $m_{\tilde{x}}=2^{-e_{\tilde{x}}}\tilde{x}-1$,  the above equations yield
\[
\tilde{y}_{00}=2^{e_{\tilde{y}_{00}}}(\tilde{\alpha}\cdot \tilde{x}+\tilde{\beta}),\]
where:
\[
\tilde{\alpha}=-2^{-(1+e_{\tilde{x}})},\quad \tilde{\beta}=\frac{1}{4}t-\frac{1}{2}e_{\tilde{x}}-e_{\tilde{y}_{00}}+\frac{1}{2},
\]
which means that $\tilde{y}_{00}$  is a piecewise linear function  of  $\tilde{x}$:
\begin{equation}
\tilde{y}_{00}(\tilde{x},t)=\left\{ 
\begin{array}{rcll}
\tilde{y}_{0}^I(\tilde{x},t)&=&-2^{-2}(\tilde{x}-2^{-1}t-3)&\text{for $\tilde{x}\in \langle 1,2)=\tilde{A}^I$}\\
\tilde{y}_{0}^{II}(\tilde{x},t)&=&-2^{-3}(\tilde{x}-t-4)&\text{for $\tilde{x}\in\langle 2,t\rangle=\tilde{A}^{II}$}\\
\tilde{y}_{0}^{III}(\tilde{x},t)&=&-2^{-4}(\tilde{x}-t-8)&\text{for $\tilde{x}\in(t,4)=\tilde{A}^{III}$}
\end{array}
\right..\label{r3_13}
\end{equation}

\ods

\begin{cor}
${\tilde y}_0 (\tilde x)$ can be approximated by piece-wise linear function  
\be
{\tilde y}_{00} (\tilde x) : = {\tilde y}_0 (\tilde x, t_1) , 
\ee
where  $t_1 = 2 + 4 m_R + 2 N_m^{-1}$  with a good accuracy $(2 N_m)^{-1} \approx  5.96 \cdot 10^{-8}$. 
\end{cor}

\begin{figure}
\begin{center}
\includegraphics[width=12cm]{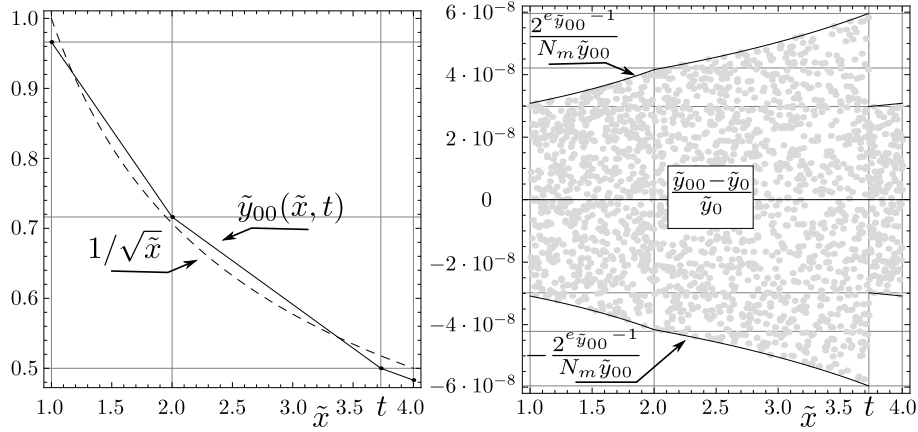}
\end{center}
\caption{Left: function $1/\sqrt{\tilde{x}}\,$  and its zeroth approximation $\tilde{y}_{00}(\tilde{x},t)$ given by (\ref{r3_13}). Right: relative error of the zeroth approximation $\tilde{y}_{00}(\tilde{x},t)$ for $2000$  random values of $\tilde{x}$.}
\label{pic2}
\end{figure}

This function  is presented on Fig.~\ref{pic2}  for a particular  value of  $m_R$: 
$$
m_R=(R-190 N_m)/N_m= 3630127/N_m\simeq 0.4327449,
$$
known from the literature. The right part of  Fig.~\ref{pic2} shows a very small relative error $(\tilde{y}_{00}-\tilde{y}_0)/\tilde{y}_0$, which confirms the validity and accuracy of the approximation (\ref{r3_10}).

In order to improve the accuracy, the zeroth approximation ($\tilde{y}_{00}$)  is corrected twice using the Newton-Raphson method (lines \textit{6} and \textit{7} in the \textit{InvSqrt} code):
\begin{align*}
\tilde{y}_{01}& = \tilde{y}_{00}-f(\tilde{y}_{00})/f'(\tilde{y}_{00}),\\
\tilde{y}_{02}& = \tilde{y}_{01}-f(\tilde{y}_{01})/f'(\tilde{y}_{01}) ,
\end{align*}
where  $f(y)=y^{-2}-x$. Therefore:
\begin{align}
\tilde{y}_{01}(\tilde{x},t)& =2^{-1} \tilde{y}_{00}(\tilde{x},t)(3-\tilde{y}_{00}^2(\tilde{x},t) \,\tilde{x}),\label{r3_14a}\\
\tilde{y}_{02}(\tilde{x},t)& =2^{-1} \tilde{y}_{01}(\tilde{x},t)(3-\tilde{y}_{01}^2(\tilde{x},t)\, \tilde{x}).\label{r3_14b}
\end{align}

In this section we gave a theoretical explanation of the \textit{InvSqrt} code. In the original form of the code the magic constant $R$ was guessed. Our interpretation gives us a natural possibility to treat $R$ as a free parameter. In next section we will find its optimal values, minimizing errors.

\section{Minimization of the relative error}
Approximations of the inverse square root presented in the previous section depend on the parameter $t$  directly related to the magic constant.   The value of this parameter can be estimated  by analysing the relative error of  $\tilde{y}_{0k}(\tilde{x},t)$ with respect to $\sqrt{\tilde{x}}$:
\begin{equation}
\tilde{\delta}_{k}(\tilde{x},t)=\sqrt{\tilde{x}}\tilde{y}_{0k}(\tilde{x},t)-1, \quad\text{where: $k\in\{0,1,2\}$}  .    \label{r4_1}
\end{equation}
As the best estimation we consider  $t=t_k^{(r)}$  minimizing  the relative error  $\tilde{\delta}_{k}(\tilde{x},t)$:
\begin{equation}
\forall_{t\neq t_k^{(r)}}\max_{\tilde{x}\in\tilde{A}}|\tilde{\delta}_{k}(\tilde{x},t_k^{(r)})|<\max_{\tilde{x}\in\tilde{A}}| \tilde{\delta}_{k}(\tilde{x},t)|\quad \text{where: $\tilde{A}=\tilde{A}^I\cup\tilde{A}^{II}\cup\tilde{A}^{III}$}.\label{r4_2}
\end{equation}

\subsection{Zeroth approximation}
\begin{figure}
\begin{center}
\includegraphics[width=11cm]{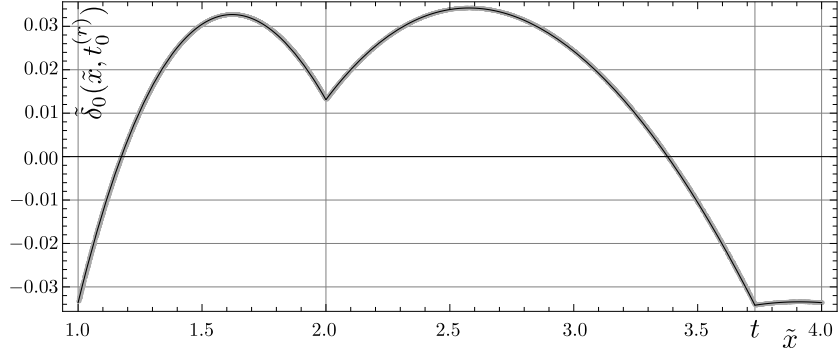}
\end{center}
\caption{Relative error for zeroth approximation of the inverse square root.  Grey points were generated by the function  \textit{InvSqrt} without two lines ($6$ and $7$) of code and with $R=R_0^{(r)}$,  for $4000$ random values  $x\in\langle 2^{-126},2^{128})$. }
\label{pic3}
\end{figure}

In order to determine $t_0^{(r)}$  we have to find extrema of  $\tilde{\delta}_0(\tilde{x},t)$ with respect to $\tilde{x}$, to identify maxima, and to compare them with boundary values  $\tilde{\delta}_{0}(1,t)=\tilde{\delta}^I_{0}(1,t)$, $\tilde{\delta}_{0}(2,t)=\tilde{\delta}^I_{0}(2,t)=\tilde{\delta}^{II}_{0}(2,t)$, $\tilde{\delta}_{0}(t,t)=\tilde{\delta}^{II}_{0}(t,t)=\tilde{\delta}^{III}_{0}(t,t)$, $\tilde{\delta}_{0}(4,t)=\tilde{\delta}^{III}_{0}(4,t)$  at the ends of the considered  intervals  $\tilde{A}^I$, $\tilde{A}^{II}$, $\tilde{A}^{III}$.  Equating to zero derivatives of $\tilde{\delta}_0^I(\tilde{x},t)$, $\tilde{\delta}_0^{II}(\tilde{x},t)$, $\tilde{\delta}_0^{III}(\tilde{x},t)$:
\begin{align}
0&=\partial_{\tilde{x}}\tilde{\delta}_0^{I}(\tilde{x},t)=2^{-3}x^{-1/2}(3-3x+2^{-1}t),\nonumber\\
0&=\partial_{\tilde{x}}\tilde{\delta}_0^{II}(\tilde{x},t)=2^{-2}x^{-1/2}(1-3\cdot 2^{-2}x+2^{-2}t),\nonumber\\
0&=\partial_{\tilde{x}}\tilde{\delta}_0^{III}(\tilde{x},t)=2^{-2}x^{-1/2}(1-3\cdot 2^{-3}x+2^{-3}t),\label{r4_3}
\end{align}
we find local extrema:
\begin{equation}
\tilde{x}_0^{I}=(6+t)/6,\quad \tilde{x}_0^{II}=(4+t)/3,\quad \tilde{x}_0^{III}=(8+t)/3.,\label{r4_4}
\end{equation}
We easily verify that $\tilde{\delta}_0^I(\tilde{x},t)$, $\tilde{\delta}_0^{II}(\tilde{x},t)$, $\tilde{\delta}_0^{III}(\tilde{x},t)$  are concave functions:
\[
\partial^2_{\tilde{x}}\tilde{\delta}_0^{K}(\tilde{x},t)<0, \quad\text{for: $K \in\{I,II,III\}$} 
\]
which means that  we have local maxima at $\tilde{x}_0^{K}$ (where $K \in\{I,II,III\}$).  One of them is negative: 
\[
\tilde{\delta}_{0m}^{III}(t)=\tilde{\delta}_0(\tilde{x}_0^{III},t)=-2^{-1}+2^{-3}t  < 0  \quad \text{for $t\in(2,4)$} .
\]
The other maxima, given by
\begin{align}
\tilde{\delta}_{0m}^{I}(t)&=\tilde{\delta}_0(x_0^I,t)=-1+2^{-1}(1+t/6)^{3/2},\nonumber\\ 
\tilde{\delta}_{0m}^{II}(t)&=\tilde{\delta}_0(x_0^{II},t)=-1+2\cdot 3^{-3/2}(1+t/4)^{3/2},\label{r4_5}
\end{align}
are increasing functions of $t$, satisfying 
\begin{align}
\tilde{\delta}_{0m}^{II}(t)& <\tilde{\delta}_{0m}^{I}(t)\quad \wedge \quad \tilde{\delta}_{0m}^{I}(t)\leq 0,&\quad &\text{for $t\in(2,3\cdot 2^{5/3}-6\rangle$},\label{r4_6}\\
\tilde{\delta}_{0m}^{II}(t)& <\tilde{\delta}_{0m}^{I}(t)\quad \wedge \quad \tilde{\delta}_{0m}^{I}(t)>0,&\quad &\text{for $t\in(3\cdot 2^{5/3}-6,2^{5/3}+2^{4/3}-2)$},\label{r4_7}\\
\tilde{\delta}_{0m}^{II}(t)& \geq \tilde{\delta}_{0m}^{I}(t)\quad \wedge \quad \tilde{\delta}_{0m}^{I}(t)>0,&\quad &\text{for $t\in\langle2^{5/3}+2^{4/3}-2,4)$}.\label{r4_8}
\end{align}
Because functions $\tilde{\delta}_0^{I}(\tilde{x},t)$, $\tilde{\delta}_0^{II}(\tilde{x},t)$, $\tilde{\delta}_0^{III}(\tilde{x},t)$  are concave, their minimal   values with respect to $\tilde{x}$  are assumed  at  boundaries of the intervals $\tilde{A}^{I}$, $\tilde{A}^{II}$, $\tilde{A}^{III}$. It turns out that the global minimum is described by the following function:
\begin{equation}
\tilde{\delta}_0^{II}(t,t)=\tilde{\delta}_0^{III}(t,t)=-1+2^{-1}\sqrt{t} ,\label{r4_9}
\end{equation}
which is increasing and negative for $t\in(2,4)$.  Therefore, taking into account  (\ref{r4_6}), (\ref{r4_7}) and (\ref{r4_8}),  the condition (\ref{r4_2}) reduces to 
\begin{align}
\tilde{\delta}_{0m}^{I}(t_0^{(r)})&=|\tilde{\delta}_0^{II}(t_0^{(r)},t_0^{(r)})|&\quad &\text{for \ \ $t_0^{(r)}\in(2,-2+2^{4/3}+2^{5/3})$},\label{r4_10}\\
\tilde{\delta}_{0m}^{II}(t_0^{(r)})&=|\tilde{\delta}_0^{II}(t_0^{(r)},t_0^{(r)})| &\quad &\text{for \  \  $t_0^{(r)}\in\langle-2+2^{4/3}+2^{5/3},4)$}.\label{r4_11}
\end{align}
The right answer results from equation (\ref{r4_11}):
\begin{equation}
t_0^{(r)}\simeq 3.7309796.\label{r4_12}
\end{equation}
Thus we obtain an estimation minimizing maximal relative error of zeroth approximation: 
\begin{equation}
\delta_{0\max}=\max_{\tilde{x}\in\tilde{A}}|\delta_{0}(\tilde{x},t_0^{(r)})|=|\delta_{0}(t_0^{(r)},t_0^{(r)})|\simeq 0.03421281  ,   \label{r4_14}
\end{equation}
and the magic constant $R_0^{(r)}$:
\begin{align}
R_0^{(r)}&=N_m(e_R+B)+\lfloor 2^{-2}N_m(t_0^{(r)}-2)-2^{-1}\rceil=\nonumber\\
&=1597465647=0x5F37642F\,.\label{r4_13}
\end{align}
The resulting relative error is presented at Fig.\ref{pic3}.

\subsection{Newton-Raphson corrections}

The relative error can be reduced by Newton-Raphson corrections  (\ref{r3_14a}) and (\ref{r3_14b}).  Substituting $\tilde{y}_{0k}(\tilde{x},t)=(1+\tilde{\delta}_k(\tilde{x},t))/\sqrt{\tilde{x}}$  \  we rewrite them as
\begin{equation}
\tilde{\delta}_k(\tilde{x},t)=-\frac{1}{2}\tilde{\delta}_{k-1}^2(3+\tilde{\delta}_{k-1}(\tilde{x},t)),\quad \text{where: $k\in\{1,2\}$}.\label{r4_15}
\end{equation}
The quadratic dependence on $\tilde{\delta}_{k-1}$ implies a fast convergence of the Newton-Raphson iterations. 
Note that obtained functions $\tilde{\delta}_{k}$  are non-positive. Their derivatives with respect to $\tilde{x}$ can be easily calculated
\begin{align}
\partial_{\tilde{x}}\tilde{\delta}_k(\tilde{x},t)&= -\frac{3}{2}\tilde{\delta}_{k-1}(\tilde{x},t)(2+\tilde{\delta}_{k-1}(\tilde{x},t)) \partial_{\tilde{x}}\tilde{\delta}_{k-1}(\tilde{x},t)\nonumber\\
\partial^2_{\tilde{x}}\tilde{\delta}_k(\tilde{x},t)&= -\frac{3}{2}\tilde{\delta}_{k-1}(\tilde{x},t)(2+\tilde{\delta}_{k-1}(\tilde{x},t)) \partial^2_{\tilde{x}}\tilde{\delta}_{k-1}(\tilde{x},t)+\nonumber\\
&-3(\tilde{\delta}_{k-1}(\tilde{x},t)+1)[\partial_{\tilde{x}}\tilde{\delta}_{k-1}(\tilde{x},t)]^2\nonumber
\end{align}
One can easily see that that extremes of $\tilde{\delta}_k$ can be determined by studying extremes  and zeros of $\tilde{\delta}_{k-1}$. 
\begin{itemize}
\item Local maxima of $\tilde{\delta}_{k}(\tilde{x},t)$  correspond to negative local minima of $\tilde{\delta}_{k-1}(\tilde{x},t)$  or for zeros of   $\tilde{\delta}_{k-1}(\tilde{x},t)$.  However,  a local maximum of  a non-positive function can not be a candidate for a global maximum of its modulus (compare  (\ref{r4_2})).
\item Local minima of $\tilde{\delta}_{k}(\tilde{x},t)$  correspond to  positive maxima and negative minima of $\tilde{\delta}_{k-1}(\tilde{x},t)$, which means that they are given by $\tilde{\delta}_{0m}^I(t)$, $\tilde{\delta}_{0m}^{II}(t)$, see (\ref{r4_5}). 
\end{itemize}   
Then, we compute 
\be
\frac{\partial\tilde{\delta}_{k}(\tilde{x},t)}{\partial\tilde{\delta}_{k-1}(\tilde{x},t)} =-\frac{3}{2}\tilde{\delta}_{k-1}(\tilde{x},t)(2+\tilde{\delta}_{k-1}(\tilde{x},t))  ,
\ee 
which implies that $\tilde{\delta}_{k}(\tilde{x},t)$ is an increasing function of $\tilde{\delta}_{k-1}(\tilde{x},t)$ for $\tilde{\delta}_{k-1}(\tilde{x},t) < 0$ and a decreasing function of $\tilde{\delta}_{k-1}(\tilde{x},t)$ for $\tilde{\delta}_{k-1}(\tilde{x},t) > 0$.  It means that there only two candidates for the minimum of $\tilde{\delta}_{1}(\tilde{x},t)$:  one corresponding to the  minimal negative value of   $\tilde{\delta}_{k-1}(\tilde{x},t)$ and one corresponding to the maximal positive value of  $\tilde{\delta}_{k-1}(\tilde{x},t)$. The smallest value of  $\tilde{\delta}_{k}(\tilde{x},t)$ evaluated at boundaries of regions $\tilde{A}^I$, $\tilde{A}^{II}$ and $\tilde{A}^{III}$   still is assumed at $\tilde{x}=t$. 

In the case $k=1$ (the first Newton-Raphston correction) we have negative minima ($\tilde{\delta}_{1m}^{I}(t)$ and $\tilde{\delta}_{1m}^{II}(t)$)  corresponding to positive maxima of  $\tilde{\delta}_{0}(\tilde{x},t)$.   The minima  are decreasing functions of $t$ and satisfy conditions
\begin{align}
|\tilde{\delta}_{1m}^{II}(t)|& <|\tilde{\delta}_{1m}^{I}(t)|\quad \text{for $t\in(2,-2+2^{4/3}+2^{5/3})$},\label{r4_16}\\
|\tilde{\delta}_{1m}^{II}(t)|& \geq |\tilde{\delta}_{1m}^{I}(t)|\quad \text{for $t\in\langle-2+2^{4/3}+2^{5/3},4)$} . \label{r4_17}
\end{align}
In the case $k=2$ we get minima at the same locations as for $k=1$ and with the same monotonicity with respect to $t$. 
All this leads to the conclusion that the condition  \rf{r4_2}  for the first and second correction will be satisfied for the same value $t=t_1^{(r)}=t_2^{(r)}$  (smaller than  $t_0^{(r)}$):
\begin{equation}
t_1^{(r)}=t_2^{(r)}\simeq 3.7298003\,,\label{r4_18}
\end{equation}
where  $t_1^{(r)}$ is a solution to the equation 
\begin{equation}
\tilde{\delta}_1(t_1^{(r)},t_1^{(r)})= \tilde{\delta}_1(\underbrace{4/3+t_1^{(r)}/3}_{\tilde{x}_0^{II}},t_1^{(r)})\label{r4_19}.
\end{equation}
Thus we obtained a new estimation of the magic constant:
\begin{align}
R_1^{(r)}=R_2^{(r)}&=N_m(e_R+B)+\lfloor 2^{-2}N_m(t_1^{(r)}-2)-2^{-1}\rceil=\nonumber\\
&=1597463174=0x5F375A86\,\label{r4_20}
\end{align}
with corresponding maximal relative errors:
\begin{equation}
\ba{l}
\tilde{\delta}_{1\max}=|\tilde{\delta}_1(t_1^{(r)},t_1^{(r)})|\simeq 1.75118\cdot 10^{-3}\,,  \\[2ex]
\tilde{\delta}_{2\max}=|\tilde{\delta}_2(t_1^{(r)},t_1^{(r)})|\simeq 4.60\cdot 10^{-6}\,.
\ea
\end{equation}

\begin{figure}
\begin{center}
\includegraphics[width=11cm]{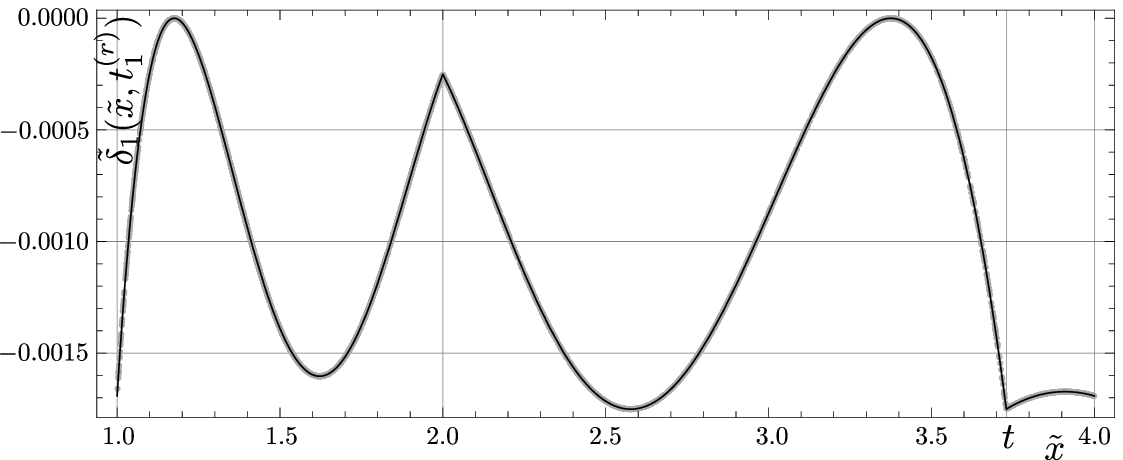}
\end{center}
\caption{Relative error of the first Newton-Raphson correction of the inverse square root approximation. Grey points were generated by the fuction \textit{InvSqrt} without line \textit{7} of the code and with  $R=R_1^{(r)}$, for $4000$ random values  $x\in\langle 2^{-126},2^{128})$.}
\label{pic4}
\end{figure}

\begin{figure}
\begin{center}
\includegraphics[width=11cm]{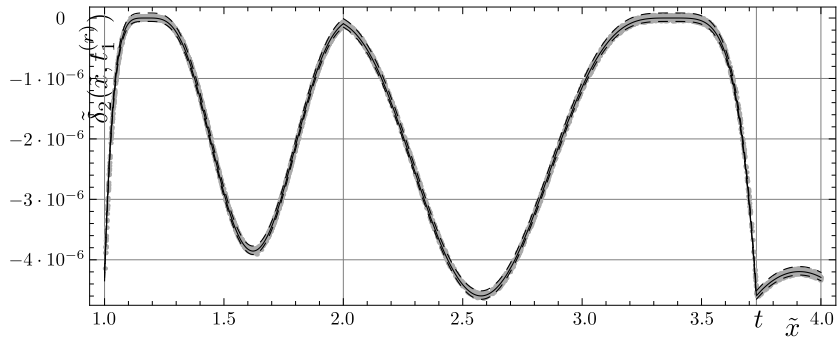}
\end{center}
\caption{Relative error of the second  Newton-Raphson correction of the inverse square root approximation. Grey points were generated by the function  \textit{InvSqrt} z $R=R_1^{(r)}=R_2^{(r)}$, for $4000$  random values  $x\in\langle 2^{-126},2^{128})$. The visible blur is a consequence of round-off errors.}
\label{pic5}
\end{figure}

\section{Minimization of the absolute error}

Similarly as in the previous section we will derive optimal values of the magic constant minimizing the absolute error of approximations $\tilde{y}_{00}(\tilde{x},t)$,  $\tilde{y}_{01}(\tilde{x},t)$  and  $\tilde{y}_{02}(\tilde{x},t)$, i.e., by minimizing 
\begin{equation}
\tilde\Delta_{k}(\tilde{x},t)=\tilde{y}_{0k}(\tilde{x},t)-1/\sqrt{\tilde{x}}, \qquad  \quad   k\in\{0,1,2\}  .         \label{r5_1}
\end{equation}
In other words, we will find $t_0^{(a)}$, $t_1^{(a)}$ and $t_2^{(a)}$  such that 
\begin{equation}
\forall_{t\neq t_k^{(a)}}\max_{\tilde{x}\in\tilde{A}}|\tilde\Delta_{k}(\tilde{x},t_k^{(a)})|<\max_{\tilde{x}\in\tilde{A}}| \tilde\Delta_{k}(\tilde{x},t)|            \label{r5_2}
\end{equation}
for $k=0,1,2$,  where \ $\tilde{A}=\tilde{A}^I\cup\tilde{A}^{II}\cup\tilde{A}^{III}$.

\subsection{Zeroth approximation}
In order to find  $t_0^{(a)}$ minimizing the absolute error of  $\tilde{y}_{00}(\tilde{x},t)$, we will compute and study extremes of  $\tilde\Delta_{0}(\tilde{x},t)$ inside intervals   $\tilde{A}^I$, $\tilde{A}^{II}$ and $\tilde{A}^{III}$, and compare them with values of  $\tilde\Delta_{0}(\tilde{x},t)$  at the ends of the intervals.  First, we compute the boundary values:
\be \ba{l}
\dis   \tilde{\Delta}_{0}(1,t) =   \frac{t}{8} - \frac{1}{2} ,  \\[2ex]
\dis  \tilde{\Delta}_{0}(2,t) =   \frac{t}{8} - \frac{2 \sqrt{2} - 1}{4} , \\[2ex]
\dis   \tilde{\Delta}_{0}(t,t) =    \frac{1}{2} - \frac{1}{\sqrt{t}}  , \\[2ex]
\dis  \tilde{\Delta}_{0}(4,t) =     \frac{t}{16} - \frac{1}{4} .  \\[2ex]
\ea \ee
All of them are negative (for $t<4$). 
Derivatives of error functions 
 ($\tilde{\Delta}_0^I(\tilde{x},t)$, $\tilde{\Delta}_0^{II}(\tilde{x},t)$  and  $\tilde{\Delta}_0^{III}(\tilde{x},t)$)   are given by:
\be  \ba{l}   \label{r5_3}
\dis  \partial_{\tilde{x}}\tilde{\Delta}_0^{I}(\tilde{x},t)=2^{-1} \tilde{x}^{-3/2}-2^{-2},  \\[1ex]
\dis \partial_{\tilde{x}}\tilde{\Delta}_0^{II}(\tilde{x},t)=2^{-1} \tilde{x}^{-3/2}-2^{-3},  \\[1ex]
\dis  \partial_{\tilde{x}}\tilde{\Delta}_0^{III}(\tilde{x},t)=2^{-1} \tilde{x}^{-3/2}-2^{-4} . 
\ea\ee
Therefore local extrema are located at
\begin{equation}   \label{minima}
\tilde{x}_{0a}^{I}=2^{2/3},\qquad \tilde{x}_{0a}^{II}=2^{4/3},\qquad \tilde{x}_{0a}^{III}=4 
\end{equation}
(the locations do not depend on $t$).  The second derivative is negative:   
\[
\partial^2_{\tilde{x}}\tilde{\Delta}_0^{K}(\tilde{x},t)=-3\cdot 2^{-2}x^{-5/2}<0   
\]
for \  $K \in\{I,II,III\}$. Therefore all these extremes are local maxima, given by  
\begin{equation}     \ba{l}    \label{r5_4}
\dis  \tilde{\Delta}_{0}(\tilde{x}_{0a}^I,t)= \tilde{y}^I_{0}(\tilde{x}_{0a}^I,t)-(\tilde{x}_{0a}^I)^{-1/2}=\frac{3}{4}-\frac{3}{2\sqrt[3]{2}}+\frac{1}{8}t  , \\[2ex] 
\dis   \tilde{\Delta}_{0}(\tilde{x}_{0a}^{II},t)= \tilde{y}^{II}_{0}(\tilde{x}_{0a}^{II},t)-(\tilde{x}_{0a}^{II})^{-1/2}=\frac{1}{2}-\frac{3 \sqrt[3]{2}}{4}+\frac{1}{8}t , 
\\[2ex] 
\dis   \tilde{\Delta}_{0}(\tilde{x}_{0a}^{III},t)= \tilde{y}^{III}_{0}(\tilde{x}_{0a}^{III},t)-(\tilde{x}_{0a}^{III})^{-1/2}=\frac{1}{16}t - \frac{1}{4}  .
\ea \end{equation}
We see that  $\tilde{\Delta}_{0}(\tilde{x}_{0a}^{III},t) < 0$ for  $t < 4$ (and negative maxima obviously are not important).   
Direct computation shows that for $t\leqslant 4$  the first value of \rf{r5_4} is the greatest, i.e.,  
\be
\max_{\tilde{x}\in\tilde{A}}\tilde{\Delta}_{0}(\tilde{x},t)=\tilde{\Delta}_{0}(\tilde{x}_{0a}^I,t)
\ee
Evaluating  $\tilde{\Delta}_{0}(\tilde{x},t)$  at the ends of the intervals $\tilde{A}^I$, $\tilde{A}^{II}$ and $\tilde{A}^{III}$,  we find the global minimum:
\begin{equation}
\min_{\tilde{x}\in\tilde{A}}\tilde{\Delta}_{0}(\tilde{x},t)=\tilde{\Delta}^I_{0}(1,t)=-\frac{1}{2}+\frac{t}{8}<0,\label{r5_5}
\end{equation}
which enables us to formulate the condition  (\ref{r5_2})  in the following form:
\begin{equation}
\max_{\tilde{x}\in\tilde{A}}\tilde{\Delta}_{0}(\tilde{x},t)= |\min_{\tilde{x}\in\tilde{A}}\tilde{\Delta}_{0}(\tilde{x},t)|\quad \Leftrightarrow\quad \frac{3}{4}-\frac{3}{2\sqrt[3]{2}}+\frac{1}{8}t=\frac{1}{2}-\frac{t}{8}. \label{r5_6}
\end{equation}   
Solving this equation we get: 
\begin{equation}
t_0^{(a)}=-1+3\cdot 2^{2/3}\simeq 3.7622,\label{r5_7}
\end{equation}
which corresponds to a magic constant $R_0^{(a)}$  given by
\begin{equation}
R_0^{(a)}=N_m(e_R+B)+\lfloor 2^{-2}N_m(t_0^{(a)}-2)-2^{-1}\rceil=1597531127=0x5F3863F7  . \label{r5_8}
\end{equation}
The resulting  maximal error of the zeroth approxmation reads 
\begin{equation}
\Delta_{0 \max}=\max_{\tilde{x}\in\tilde{A}}|\Delta_{k}(\tilde{x},t_k^{(a)})|=\frac{5}{8}-\frac{3}{4\sqrt[3]{2}}\simeq 0.0297246.\label{r5_9}
\end{equation}
\begin{figure}[h]
\begin{center}
\includegraphics[width=11cm]{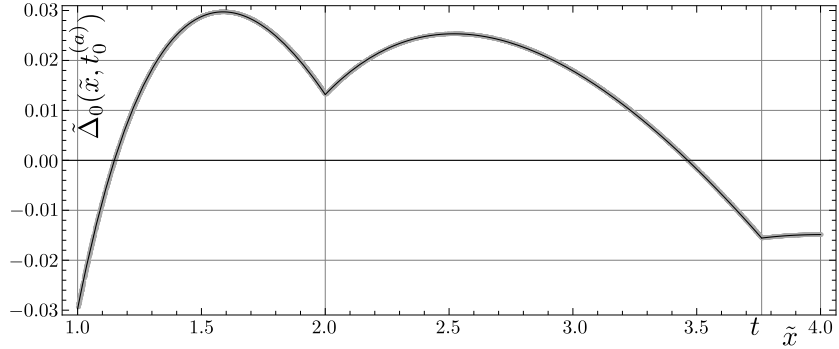}
\end{center}
\caption{Absolute error of zeroth approximation of the inverse square root. Grey points were generated by the function \textit{InvSqrt} without two lines ($6$ and $7$) of the code and with $R=R_0^{(a)}$, for $4000$ random values $x \in\langle 1,4)$.}
\label{pic6}
\end{figure}

\subsection{Newton-Raphson corrections}
\begin{figure}
\begin{center}
\includegraphics[width=11cm]{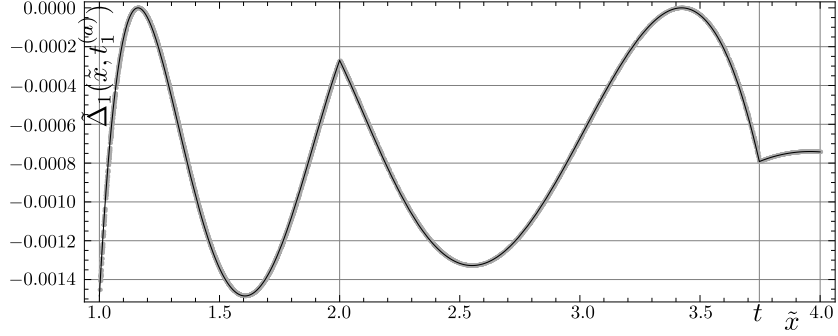}
\end{center}
\caption{Absolute error of the first  Newton-Raphson correction of the inverse square root approximation. Grey points were generated by the function  \textit{InvSqrt}  without the line $7$  of the \textit{InvSqrt} code and with  $R=R_1^{(a)}$, for $4000$  random values  $\tilde x\in\langle 1,4)$.} 
\label{pic7}
\end{figure}
The absolute errror after Newton-Raphson corrections is a non-positive function, similarly as the relative error. This function reaches its maximal value equal to zero in intervals $\tilde{A}^I$ and $\tilde{A}^{II}$ (which corresponds to zeros of  $\tilde{\Delta}_0^I$ i $\tilde{\Delta}_0^{II}$) and has a negative maximum in the interval  $\tilde{A}^{III}$. The  other extremes  (minima) are decreasing functions of the parameter $t$. They are  located at $x$ defined by the following equations:  
\begin{align}
0=&-\frac{75}{128}-\frac{27t}{256}-\frac{9t^2}{512}-\frac{t^3}{1024}+\frac{1}{2x^{3/2}} +\frac{27x}{64}+\frac{9tx}{64}+\frac{3t^2x}{256}-\frac{27x^2}{128}+\nonumber\\
&-\frac{9tx^2}{256}+\frac{x^3}{32},\;\;\qquad\qquad\qquad\qquad\text{for $x\in\tilde{A}^I$},\label{r5_10a}
\end{align}
\begin{align}
0=&-\frac{1}{4}-\frac{3t}{64}-\frac{3t^2}{256}-\frac{t^3}{1024}+\frac{1}{2x^{3/2}} +\frac{3x}{32}+\frac{3tx}{64}+\frac{3t^2x}{512}-\frac{9x^2}{256}+\nonumber\\
&-\frac{9tx^2}{1024}+\frac{x^3}{256},\qquad\qquad\qquad\qquad\text{for $x\in\tilde{A}^{II}$}\label{r5_10b}.
\end{align}
The condition (\ref{r5_2})  reduces to the equality of  the local minimum (located in $\tilde{A}^I$)  and  $\tilde{\Delta}_1(1,t)$ (this is an increasing function of $t$).  The equality is obtained for $t=t_{1}^{(a)}$, where 
\begin{equation}
t_1^{(a)}\simeq 3.74699138 ,   \label{r5_11}
\end{equation}
The corresponding maximal error and a magic constant are given by
\begin{equation}
\Delta_{1 \max}\simeq 0.001484497,\label{r5_12}
\end{equation}
\begin{equation}
R_1^{(a)}=1597499226=0x5F37E75A\,.\label{r5_13}
\end{equation}
In the case of the second  Newton-Raphson correction the minimization of errors is obtained similarly, by equating the local minimum $\tilde{\Delta}_2^{I}(\tilde{x},t)$   with   $\tilde{\Delta}_2^I(1,t)$.  Hence we get  another value a magic constant:
\begin{equation}
R_2^{(a)}=1597484501=0x5F37ADD5,\label{r5_14}
\end{equation}
corresponding to 
\begin{equation}
t_2^{(a)}\simeq 3.73996986,\quad \Delta_{2 \max}\simeq 3.684\cdot 10^{-6}.\label{r5_15}
\end{equation}
\begin{figure}
\begin{center}
\includegraphics[width=11cm]{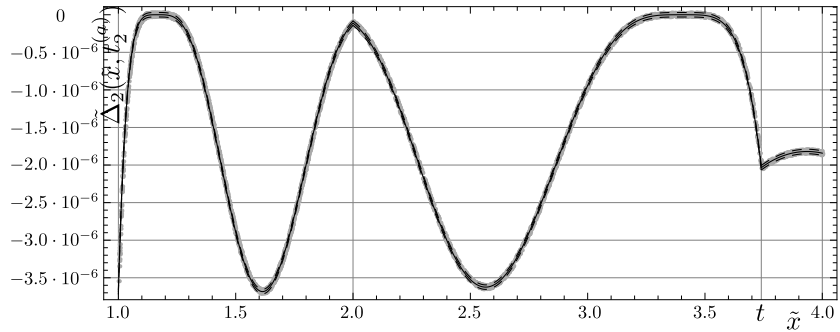}
\end{center}
\caption{Absolute error of the second  Newton-Raphson correction of the inverse square root approximation. Grey points were generated by the function  \textit{InvSqrt}  of the \textit{InvSqrt} code and with  $R=R_2^{(a)}$, for $4000$  random values  $\tilde x\in\langle 1,4)$.}
\label{pic8}
\end{figure}

\section{Conclusions}

In this paper we have presented a theoretical interpretation of the \textit{InvSqrt} code, giving a precise meaning to two values of the magic constant exisiting in the literature and adding two more values of the magic constant. Using this magic constant we have conducted error analysis for Newton-Raphson Method and proved that error bounds for single precision  computation are acceptable. The magic constant can be easily incorporated in existing floating point multiplier or floating point multiply-add fused and one need to replace LUT with the magic constant.

\end{document}